\documentclass[a4paper,oneside,fleqn,draft]{article}
\usepackage{amsmath,amssymb,lscape,sectsty,overcite}
\newcommand{\nc}{\newcommand}
\nc{\cc}{\mathrm{c.c.}}
\nc{\zet}{z}
\nc{\bzet}{\bar z}
\nc{\bea}{\begin{eqnarray}}
\nc{\eea}{\end{eqnarray}}
\nc{\be}{\begin{equation}}
\nc{\beag}{\begin{eqnarray*}}
\nc{\eeag}{\end{eqnarray*}}
\nc{\ee}{\end{equation}}
\nc{\iz}{\,{\scriptstyle{\wedge}}\,}
\nc{\frec}[2]{{\textstyle{\frac{#1}{#2}}}}
\nc{\ud}{\mathrm{d}}\nc{\Tr}{\mathrm{Tr}\,}
\nc{\heq}{\,\,\hat{=}\,\,}
\nc{\nheq}{\,\,\hat{\neq}\,\,}
\nc{\const}{\mathrm{const}}
\nc{\bxi}{\bar{\xi}}
\nc{\bz}{\bar{z}}
\nc{\bzeta}{\bar{\zeta}}
\nc{\bF}{\bar{F}}
\nc{\hP}{\hat{P}}\nc{\hu}{\hat{u}}
\nc{\sgn}{\mathrm{sgn}}
\nc{\cov}{{\scriptscriptstyle |}}
\nc{\scri}{\mathcal{I}^+}
\nc{\contr}{-\!\!\!\lower-.260em\hbox{$\lrcorner$}\,}
\nc{\bes}{\begin{subequations}}
\nc{\ees}{\end{subequations}}
\nc{\alg}{\mathfrak{g}}
\allsectionsfont{\normalsize}
\begin{document}
\makeatletter \renewcommand\@biblabel[1]{$^{#1}\!\!\!$} \makeatother
\title{Symmetries of the Robinson--Trautman equation}
\author{W. Natorf and J. Tafel,\\
  Institute of Theoretical Physics,\\
  Warsaw University, ul. Ho\.za 69 00--681,\\
  Warsaw, Poland}\date{}
\maketitle

\begin{abstract}
  \noindent We study point symmetries of the Robinson--Trautman
  equation. The cases of one- and two-dimensional algebras of
  infinitesimal symmetries are discussed in detail. The corresponding
  symmetry reductions of the equation are given. Higher dimensional
  symmetries are shortly discussed.  It turns out that all known exact
  solutions of the Robinson--Trautman equation are symmetric.
\end{abstract}

\section{Introduction}
In 1960 Robinson and Trautman introduced \cite{RT} a class of
space-times admitting a diverging shear-- and twist--free congruence
of null geodesics. Such space-times, if asymptotically Minkowskian,
are believed to describe gravitational radiation outgoing from
spatially bounded sources. Recent numerical results suggest that the
Robinson-Trautman metrics can be used to estimate the mass loss during
the final phase of the collision of two black holes \cite{MO}.  There
are also suggestions that the subclass of the so called C-metrics
\cite{KIN} (and their twisting generalizations \cite{PD}) can describe
spacetimes containing accelerating black holes
\cite{prawdy1,prawdy2,prawdy3,bp}.

In terms of standard coordinates $u,r,\xi,\bxi$, where $u$ and $r$ are
real and $\xi$ is complex, the Robinson--Trautman metrics are given
by\cite{KR} 
\be\label{RTmetric} g = 2\ud u(H \ud u + \ud r) - 2 r^2
P^{-2}\ud\xi\ud\bxi\ .\ee 
The function $P$ is independent of $r$.
Vacuum Einstein equations imply \be\label{H} H = -r \partial_u \ln P -
m(u)/r + P^2\partial_{\xi}
\partial_{\bxi}\ln P\ee
and a fourth order equation for $P$, referred to as the
Robinson--Trautman equation:
\be\label{RTeq}P^2\partial_{\xi}\partial_{\bxi}
(P^2\partial_{\xi}\partial_{\bxi}\ln P)+3m
\partial_u (\ln P)-\partial_u m=0\ .\ee
Using coordinate freedom $m$ can be transformed to the value
$\pm 1$ or $0$. The Gaussian curvature $K$ of surfaces of constant $u$
and $r$ is given by
\be\label{K} K= 2P^2\partial_{\xi}\partial_{\bxi}\ln P\ .\ee

Existence of asymptotically flat Robinson--Trautman metrics has been
recently examined \cite{RE,SI,CH}. Also, their large $u$ asymptotic
behaviour is known \cite{FN,MF}. These results, however, do not give
any hint of how to look for explicit solutions. Only a few of them are
known since 1960, none of them being asymptotically flat except for
the cases of the Minkowski and the Schwarzschild metrics.

Assuming asymptotic flatness of metrics \eqref{RTmetric}, the Bondi
energy \cite{HB,BB}, the Bondi mass aspect and the news function were
found in terms of $P$ \cite{SI,TD,TB,CM}.  Asymptotic flatness of
\eqref{RTmetric} with positive $m$ follows from the assumption that in
the gauge $m=1$ the function $\hP=P/(1+\frec{1}{2}\xi\bxi)$ is
positive and regular on $R\times S_2$, $\xi$ being interpreted as a
complex stereographic coordinate on $S_2$. 

The subclass of metrics with $m=0$ is characterized by the fact that
$K$ is a solution of the Laplace equation, so $K$ must be either
constant or singular. No nontrivial asymptotically flat
metric exists in this subclass.

In the present paper we consider symmetries of the Robinson-Trautman
equation and classify the conjugacy classes of one- and
two-dimensional symmetry algebras. We find symmetry reductions of equation
\eqref{RTeq} for solutions preserved by these algebras. Some of these
solutions might correspond to asymptotically flat metrics with a
simple dependence on the time $u$.  This research may also be of help
for studying the numerical solutions of the Robinson--Trautman
equation, providing natural ansatzes with smaller number of variables
than in generic situations.

\section{Symmetry transformations for $m\neq 0$}
Suppose $m\neq 0$ and consider equation \eqref{RTeq} in the gauge $m=1$,
\be\label{RTeqm1}
P^2\partial_{\xi}\partial_{\bxi}(P^2\partial_{\xi}\partial_{\bxi}\ln P)
+ 3  \partial_u\ln P = 0\ .\ee
It is easy to prove that \eqref{RTeqm1} is invariant with respect to
the following point transformations:
\be\begin{split}\label{symm1}
u&\mapsto u'=a^4u+b,\\
\xi &\mapsto\xi'= f(\xi),\\
P &\mapsto P'=a^{-1}|f_{,\xi}| P\ ,\end{split}
\ee
where $a\neq 0$ and $b$ are real constants and $f$ is a holomorphic
function of $\xi$ and $f_{,\xi}\neq 0$.
If \eqref{symm1} is supplemented by \be\label{r}r'=a^2 r\ ,\ee
the corresponding metric transforms as $g\mapsto a^6 g$. Thus,
\eqref{symm1} together with \eqref{r} induces a homothety of the
metric. It is an isometry when $a=1$.

Infinitesimal transformations corresponding to \eqref{symm1}
are given by the vector field 
\be\label{vf} k=(4Au+B)\partial_u+
F(\xi)\partial_{\xi}+\bF(\bxi)\partial_{\bxi}+
(\mathrm{Re}F_{,\xi}-A)P\partial_P\ ,\ee 
where $A$ and $B$ are real constants and $F$ is a holomorphic
function of $\xi$. These fields form the symmetry algebra $\alg$.

Solutions of \eqref{RTeq} invariant with respect to \eqref{vf}
have to satisfy the following linear equation
\be\label{inv}
(4Au+B)P_{,u}+FP_{,\xi}+\bF P_{,\bxi}+(A-\mathrm{Re}F_{,\xi})P=0\ .\ee

We will perform the symmetry reduction of \eqref{RTeqm1} assuming
that its solution is preserved by vector fields \eqref{vf}
forming a one- or two-dimensional subalgebra of $\alg$ (higher
dimensional subalgebras will also be shortly discussed.) First we will
classify these subalgebras under the action of pseudogroup of
transformations \eqref{symm1}. This way we will obtain
the conjugacy classes (CC)
of one- and two-dimensional subalgebras of $\alg$.

\section{Solutions with one or two symmetries}
Consider a one-dimensional algebra generated by a vector field \eqref{vf}.
Applying an appropriate transformation \eqref{symm1} we can simplify
the coefficients $A$ and $B$ and the function $F$, obtaining one
representative of each conjugacy class. For instance, if $A=0$, $B\neq
0$ and $F\neq 0$ we can scale $u$ and $\xi$ so that $B=1$ and $F=\xi$.
This leads to \be\label{przyklad}k = \partial_u +\xi\partial_{\xi}
+\bxi\partial_{\bxi}+P\partial_P\ .\ee In the same way we can
distinguish five conjugacy classes which are listed in Table 1,
together with corresponding vector $k$, form of the invariant solution
$P$ and the reduced Robinson--Trautman equation.  Whenever it
is needed, we explicitly write the definition of the new variable
$z(u,\xi)$ appearing in the invariant solution. Throughout the text
we use $x$ and $y$ to denote, respectively, the real and imaginary part of
$\xi$.

\medskip

Consider now the case of two-dimensional subalgebra $\alg_2$ of
$\alg$. We denote the basis vectors of $\alg_2$ by $k_1$ and
$k_2$. There are two nonisomorphic two-dimensional Lie
algebras such that either
\be\label{abelian}
[k_1,k_2]=0\ee
or
\be\label{nonabelian}
[k_1,k_2]=k_2\ .\ee
The Lie bracket of two fields given by \eqref{vf} reads
\be\label{komut}\begin{split}
[k_1,k_2]&=4(B_1A_2-B_2A_1)\partial_u + 
(F_1 F_{2,\xi}-F_2 F_{1,\xi})\partial_{\xi} + \\
&+(\bF_1 \bF_{2,\bxi}-\bF_2 \bF_{1,\bxi})\partial_{\bxi}
+P\,\mathrm{Re}(F_1F_{2,\xi\xi}-F_2F_{1,\xi\xi})\partial_P
\end{split}\ee
where the indices $1,2$ refer to vectors $k_1$, $k_2$ respectively.

In the Abelian case, equation \eqref{abelian} implies
\begin{subequations}
\label{abel}
\be
A_1 B_2 = A_2 B_1\ ,\label{abel:u}
\ee
\be
F_1 F_{2,\xi} = F_2 F_{1,\xi}\ ,\label{abel:xi}
\ee
\be
\mathrm{Re}(F_1 F_{2,\xi\xi} - F_2 F_{1,\xi\xi}) = 0\ .\label{abel:P}
\ee
\end{subequations}
It follows from \eqref{abel:xi} that $F_1$ is proportional to $F_2$
and \eqref{abel:P} is satisfied.

Due to \eqref{abel:u} we can assume without loss of generality
that $A_2=B_2=0$. Then $F_2\neq 0$ and
using the gauge freedom in $\xi$ we can set $F_2=i$.
Then $F_1=C=\const\in R$ follows from \eqref{abel:xi} and
the remaining freedom in choice of $k_1$. Therefore, the symmetry
generators are
\be\label{abelk1k2}
k_1=(4Au+B)\partial_u + C\partial_x-AP\partial_P,
\quad k_2=\partial_y\ .\ee
Invariance of a solution $P$ of \eqref{RTeqm1}
with respect to $k_2$ implies
\be\label{k2abelinv}
P=p(u,x)\ .\ee
Invariance with respect to $k_1$ gives
\be\label{k1abelinv}
(4 A u+B)P_{,u}+CP_{,x}=-AP\ .\ee
Depending on values of $A$, $B$ and $C$ we can distinguish five
conjugacy classes of two-dimensional Abelian subalgebras of $\alg$.
They are listed in Table 2. (Here, as well as in Tables 3 and 4
the symbol $p'$ denotes a derivative of
$p$ with respect to its argument.)

In the non-Abelian case, equation \eqref{nonabelian} gives
\begin{subequations}
\label{nonabel}
\be 4(A_2B_1-A_1B_2)=4A_2u+B_2\ ,\label{nonabel:u}\ee
\be F_1 F_{2,\xi}-F_2 F_{1,\xi}=F_2\ ,\label{nonabel:xi}\ee
\be P \mathrm{Re}\left(F_1F_{2,\xi\xi}-F_2F_{1,\xi\xi}\right)=
P\left(\mathrm{Re}F_{2,\xi} - A_2\right)\ .\label{nonabel:P}\ee
\end{subequations}
It follows from \eqref{nonabel:u} that
\be\label{nonabu}
A_2=0,\ B_2(1+4A_1)=0\ .\ee
Differentiating \eqref{nonabel:xi} and its complex conjugate
we conclude that \eqref{nonabel:P} follows from \eqref{nonabel:xi}.
Proceeding as in the Abelian case, we obtain results
summarized in Table 3.

\section{Solutions with more symmetries}
All Lie algebras of dimension greater than three contain a
three-dimensional subalgebra. Let $k_i$, $i=1,2,3$, be its generators.
In the case of Bianchi type {\it VIII} or {\it IX} it follows from
\eqref{komut} that $A_i=B_i=0$. Then the fields $k_1$, $k_2$ and $k_3$
are tangent to a two-dimensional surface and invariant solutions are
excluded.  For all other Bianchi types the algebra contains a
two-dimensional Abelian subalgebra. One can construct invariant
solutions as special cases of those described in Table 2.  After
lengthy calculations one obtains only the trivial solution $P=\const$
or $P=\const\cdot x^{3/2}$ given by Robinson and Trautman \cite{RT2}. These are
also solutions of the Robinson--Trautman equation for $m=0$. Thus,
assumption of three or more symmetries does not lead to any new
interesting solutions.

\section{Symmetry transformations for $m=0$}
Suppose now that $m=0$. In this case equation \eqref{RTeq} can
be integrated to a second order equation,
\be\label{RTeqm0} P^2\partial_{\xi}\partial_{\bxi}\ln P =
\mathrm{Re}\,\phi(u,\xi)\ , \ee 
where $\phi$ is holomorphic with respect
to $\xi$. If $\phi_{,\xi}\neq 0$, one can transform
\eqref{RTeqm0} to the equation \cite{KR}
\be\label{RTeqm0bezu}
P^2\partial_{\xi}\partial_{\bxi}\ln P = - \mathrm{Re}\,\xi\ .\ee
Substituting \be\label{podst}
P= x^{3/2}p(\xi,\bxi)\ee 
into \eqref{RTeqm0bezu}
we get
\be\label{RTeqm0lapl}
p^2(\Delta
\ln p-3/8)=-1\ .\ee 
Here $\Delta=x^2\partial_{\xi}\partial_{\bxi}$ is the Laplace operator
on a pseudosphere with the metric
\be\label{m2xi}g=\frac{\ud\xi\ud\bxi}{x^2}\ee
which can be put into the standard form
\be\label{m2}g=\frac{4\ud\zeta\ud\bzeta}{(1-\zeta\bzeta)^2}\ee
by means of the transformation $\zeta = (\xi+1)/(\xi-1).$
The operator $\Delta$ is preserved by transformations
\be\label{homo}\xi\mapsto\xi' = \frac{a \xi + i b}{i c\xi+d}\ ,
\quad a,b,c,d\in R\ee
corresponding to an action of SL(2,$R$) on the pseudosphere.

It follows from \eqref{homo} that infinitesimal transformations
are generated by vector fields of the form
\be\label{genm0}
k = (iA\xi^2 +B\xi+iC)\partial_{\xi}+\cc\ ,\ee
where $A$, $B$ and $C$ are real constants. Using symmetry
transformations we can distinguish three 
conjugacy classes of one-dimensional subalgebras
of the symmetry algebra for $m=0$ (Table 4).

Given a solution of \eqref{RTeqm0bezu} we can apply the transformation
$\xi'=f(u,\xi)$ to obtain a class of solutions of \eqref{RTeqm0}.

Note that the only solution of equation 
\eqref{RTeqm0lapl} invariant with respect to
two independent fields of the form \eqref{genm0} is $p=\sqrt{8/3}$ which 
gives the well known solution $P=\const\cdot x^{3/2}$ found by
Robinson and Trautman\cite{RT2}.

\newpage\landscape
\noindent Table 1. Invariant solutions with single symmetry.

\bigskip

\noindent\begin{tabular}{|c|l|l|l|}\hline
CC & $k$ & $P$ & RT equation\\\hline\hline
 &&&\\
1 & $\partial_u$ &  $p(\xi,\bxi)$  & 
$ p^2 \partial_{\xi}\partial_{\bxi}\ln p = \mathrm{Re}\,\xi.$\\
 &&&\\
2 & $4u\partial_u-P\partial_P$ &  $u^{-1/4}p(\xi,\bxi)$  & 
$ p^2\partial_{\xi}\partial_{\bxi}(p^2\partial_{\xi}\partial_{\bxi}
\ln p) = \frac{3}{4}.$ \\
&&&\\
3 & $\partial_u-\xi\partial_{\xi}-\bxi\partial_{\bxi}-P\partial_P$ & 
$e^u p(z,\bzet)$ & 
$ p^2 \partial_{\zet}\partial_{\bzet}(
p^2 \partial_{\zet}\partial_{\bzet}\ln p)-3(\zet\partial_{\zet}
+\bzet\partial_{\bzet})\ln p +3 = 0,\quad \zet = e^{-u}\xi.$ \\
&&&\\
4 & $\partial_y$ &  $p(u,x)$  & 
$p^2 \partial_x^2(p^2 \partial_x^2 \ln p) + 3 \partial_u \ln p = 0.$\\
&&&\\
5 & $4Au\partial_u+\xi\partial_{\xi}+\bxi\partial_{\bxi}+P(1-A)\partial_P$ &
$u^{\frac{1-A}{4A}}p(z,\bzet)$  & 
$4A p^2 \partial_{\zet}\partial_{\bzet}(
p^2 \partial_{\zet}\partial_{\bzet}\ln p)
+3(A-1)+3(\zet\partial_{\zet}
+\bzet\partial_{\bzet})\ln p = 0,$\\ 
&&&$\zet=u^{-(4A)^{-1}}\xi.$\\\hline\hline
\end{tabular}

\vspace{0.3cm}

\noindent Table 2. Invariant solutions with two Abelian symmetries.

\bigskip

\noindent\begin{tabular}{|c|l|l|l|l|}\hline
CC & $k_1$ & $k_2$ & $P$ &  RT equation \\\hline\hline

&&&&\\

A.1 & $\partial_u$ & $\partial_y$ &  $p(x)$ &
$p^2(\ln p)'' = x.$ \\

&&&&\\

A.2 & $\partial_x$ & $\partial_y$ & $p(u)$  & 
$p'=0.$ \\

&&&&\\

A.3 & $\partial_u +\partial_x$ & $\partial_y$ & $p(u-x)$  & 
$p^2(p^2(\ln p)'')''+3(\ln p)'=0.$ \\

&&&&\\

A.4 & $4u\partial_u-P\partial_P$ & $\partial_y$ & $u^{-1/4}p(x)$  & 
$ p^2(p^2(\ln p)'')''=\frac{3}{4}.$ \\

&&&&\\

A.5 & 
$4Au\partial_u+\partial_x-AP\partial_P$ & $\partial_y$ &
 $u^{-1/4}p((4A)^{-1}\ln u -x)$ &
$ p^2(p^2(\ln p)'')''+\frac{3}{4A}(\ln p)'=\frac{3}{4}.$\\\hline\hline
\end{tabular}

\newpage

\noindent Table 3. Invariant solutions with two non-Abelian symmetries.

\bigskip

\noindent\begin{tabular}{|c|l|l|l|l|}\hline
CC & $k_1$ & $k_2$ & $P$ & RT equation \\\hline\hline
&&&&\\

NA.1 & $\epsilon\partial_u-x\partial_x-y\partial_y-P\partial_P$,
 &
$\partial_y$ & $ xp(u+\epsilon\ln x)$ & 
$\epsilon[p^3(p''''-2p'''-p''+2p')-$ \\
 &$\epsilon=0$ or $\epsilon=1$&&& 
$-p^2(p''^2+2p'p''+p'^2)]+3(\ln p)'=0.$\\

&&&&\\

NA.2 & $4Au\partial_u-x\partial_x-y\partial_y-(1+A)P\partial_P$ &
$\partial_y$ &  $ u^{-\frac{1+A}{4A}}p(
\underbrace{
u^{\frac{1}{4A}}x
}_{X}
)$  & 
$p^2(p^2(\ln p)'')''+3X (\ln p)'=3(1+A).$ \\

&&&&\\

NA.3 & $-u\partial_u+\partial_x +\frac{1}{4}P\partial_P$ &
$\partial_u$ &  $p(x)$  & 
$p^2(\ln p)''= x.$ \\

&&&&\\

NA.4 & 
$-u\partial_u+\partial_x +\frac{1}{4}P\partial_P$ &
$\partial_u+e^{\xi}\partial_{\xi}+e^{\bxi}\partial_{\bxi}+$ & 
$u^{-\frac{1}{4}}|1+u^{-1}e^{-\xi}|^{\frac{3}{4}}
p(z),$ &

$p^2(p^2(\ln p)'')'' + \frac{1}{4}p^4(\ln p)''
-$\\
& 
& 
$+P\,\mathrm{Re}\,e^{\xi}\partial_P$
 & $z=\frac{1+u^{-1}e^{-\xi}}{1+u^{-1}e^{-\bxi}}.$
 &  $- 36\cos z+48(\ln p)'\sin z=0.$ 
\\\hline\hline
\end{tabular}

\vspace{0.3cm}

\noindent Table 4. Invariant solutions for $m=0$.

\bigskip

\noindent \begin{tabular}{|c|l|l|l|}\hline
CC & $k$ & $P$ & RT equation\\\hline\hline
0.1 & $i(\xi^2-1)\partial_{\xi}-i(\bxi^2-1)\partial_{\bxi}$ & 
$p(r)$, $r = \left|\frac{\xi+1}{\xi-1}\right|$ &  
$\frac{1}{4}(1-r^2)^2 \frac{1}{r}
(r(\ln p)')'-\frac{3}{8}+p^{-2}=0.$
 \\

&&& \\

&&&\\
0.2 & $\xi\partial_{\xi}+\bxi\partial_{\bxi}$ &  $p(\phi)$, $\xi=re^{i\phi}$ & 
$\frac{1}{4}\cos^2
\phi(\ln p)''-\frac{3}{8}+ p^{-2}=0.$ 
  \\

&&&\\

&&&\\

0.3 & $i(\partial_{\xi}-\partial_{\bxi})$ &  $p(x)$ &
$\frac{1}{4}x^2(\ln p)''-\frac{3}{8}+ p^{-2}=0.$\\
\hline\hline
\end{tabular}
\newpage\endlandscape

\section*{Discussion}
We have examined point symmetries of the Robinson--Trautman equation
\eqref{RTeq}. Forms of solutions in the case of one or two symmetries
were given, as well as the corresponding reduced equations. Note that
all known \cite{KR} exact solutions of \eqref{RTeq} have two or three
symmetries and belong to one of the cases considered here.

Assumption of symmetry does not seem to exclude regular solutions
corresponding to asymptotically flat metrics.  The coordinates
$\xi,\bxi$ used in Tables 1--3 can differ from the hypothetical
Bondi--Sachs coordinates. For instance, in the case 4 of table 1, the
function $P$ depends on $x$ which becomes $\ln|\xi'|$ under
the transformation $\xi\mapsto \xi'=\exp(\xi)$.  To achieve regularity
of $\hP$ one should demand that $P$ is everywhere finite, positive,
and \be\label{asymp}\lim_{|\xi'|\to\infty}|\xi'|^{-1}P = a(u),\quad
\lim_{|\xi'|\to 0}|\xi'|P = b(u)\ee
for some positive functions $a$ and $b$. 
Rewriting \eqref{asymp} in terms of $\xi$ we get \be\label{asympsw}
\lim_{x\to -\infty}e^x P = a(u),
\quad
\lim_{x\to \infty}e^{-x}P = b(u)\ee
which means that $P$ behaves like $e^{|x|}f(u)$ for large $|x|$.

In the case A.3 (see Table 2) the Robinson-Trautman equation can be
solved analytically. This way one obtains the so-called C-metrics. A
possible physical interpretation of these metrics as well as their
twisting generalizations (the spinning C-metrics) was done recently
\cite{prawdy1,prawdy2,prawdy3,bp}

\section*{Acknowledgments}
J. T. would like to thank M. Grundland who participated
in the early stages of these investigations.

This work was partially supported by the Polish Committee
for Scientific Research (grant 1 PO3B 075 29).


\begin{thebibliography}{15}
\bibitem{RT} Robinson I. and Trautman A.,
{\it  Spherical Gravitational Waves,}
Phys. Rev. Lett. {\bf 4} 461 (1960)
\bibitem{MO} Moreschi O., Perez A., and Lehner L.,
{\it  Energy and angular momentum radiated for non head-on binary black hole collisions,}
 Phys. Rev. D {\bf 66} 104017 (2002)
\bibitem{KIN} Kinnersley W.,
{\it  Field of an Arbitrarily Accelerating Point Mass,}
Phys. Rev. {\bf 186}, 1335 (1969)
\bibitem{PD} Demia\'nski M., Pleba\'nski J. F.,
{\it Rotating, charged and uniformly accelerating mass
in general relativity} Ann. Phys. (USA) {\bf 98} 98 (1976)
\bibitem{prawdy1}
Pravda V. and Pravdov\'a A., {\it Boost-rotation symmetric vacuum spacetimes with spinning sources} J. Math. Phys. {\bf 43} 1536 (2002)
\bibitem{prawdy2}
Pravda V. and Pravdov\'a A.,
{\it Boost-rotation symmetric spacetimes---review}
Czech. J. Phys. {\bf 50} 333 (2000)
\bibitem{prawdy3}
Pravda V. and Pravdov\'a A.,
{\it On the spinning C-metric}
Gravitation: Following the Prague Inspiration 
(Selected essays in honour of J. Bi\v{c}\'ak), Eds. 
O. Semerak, J. Podolsk\'y and M. Zofka, World Scientific,
Singapore (2002)
\bibitem{bp} Bi\v{c}\'ak J., Pravda V.,
{\it Spinning C-metric: a spacetime with accelerating,
rotating black holes},
Phys. Rev. D {\bf 40} 1827 (1989)



\bibitem{KR} Stephani H., Kramer D., MacCallum M., Hoenselaers C., and Herlt E.,
{\it  Exact Solutions to Einstein's Field Equations, Second Edition,}
2003, Cambridge University Press.
\bibitem{RE} Rendall A.,
{\it  Existence and asymptotic properties of global solutions of Robinson--Trautman equation,}
 Class. Quantum Grav. {\bf 5} 1339 (1988)
\bibitem{SI} Singleton D.,
{\it  On global existence and convergence of vacuum Robinson--Trautman solutions,}
 Class. Quantum Grav. {\bf 7} 1333, (1990)
\bibitem{CH} Chru\'sciel P. T.,
{\it  On the global structure of Robinson--Trautman space-times,}
 Proc. R. Soc. Lond. A {\bf 436} 299 (1992) 
\bibitem{FN} Foster J., and Newman E. T.,
{\it  Note on the Robinson--Trautman Solutions,}
J. Math. Phys. {\bf 8} 189 (1966)
\bibitem{MF} Frittelli S., and Moreschi O. M.,
{\it  Study of the Robinson--Trautman metrics in the asymptotic future,}
Gen. Rel. Grav. {\bf 24} 575 (1992)
\bibitem{HB}
Bondi H., {\it Gravitational Waves in General Relativity,} Nature (London) {\bf 186}
535 (1960)
\bibitem{BB}
Bondi H.,  van der Burg M. G. J., and  Metzner A. W. K., 
{\it Gravitational Waves in General Relativity. VII. Waves from Axi-Symmetric
Isolated Systems,}
Proc. Roy. Soc. London {\bf A269} 21
(1962)
\bibitem{TD} Tod P., 
{\it Some Examples of Penrose's Quasi Local Mass Construction,}
Proc. Roy. Soc. Lond. A {\bf 388} 457, (1983)

\bibitem{TB} Tafel J.,
{\it  Bondi mass in terms of the Penrose conformal factor,}
Class. Quantum Grav. {\bf 17} 4397 (2000)

\bibitem{CM} Cornish F. H. J., and Micklewright B.,
{\it  The news function for Robinson--Trautman radiating metrics}
Class. Quantum Grav. {\bf 16} 611 (1999)

\bibitem{RT2} Robinson I., and Trautman A.,
{\it  Some spherical gravitational waves in general relativity,}
Proc. Roy. Soc. London A {\bf 265} 463 (1962)

\end{thebibliography}
\end{document}